\documentstyle[sprocl,epsfig]{article}
\begin{document}
\baselineskip=24ptplus.5ptminus.2pt
\setcounter{page}{1}
\vspace*{0.5 in}
\large

\begin{center}
Quark-Meson Coupling Model for a Nucleon \\
\end{center}

\normalsize

\begin{center}
S. W. Hong$^{\dagger }$, B. K. Jennings$^{\ast}$
\par
$^{\dagger}$
Department of Physics and Institute of Basic Science, \\
Sungkyunkwan University, Suwon 440-746, Korea \\
$^{\ast}$
TRIUMF, 4004 Wesbrook Mall, Vancouver,\\ British Columbia, Canada V6T 2A3 \\
\end{center}
\vspace{2ex}

\begin{center}
ABSTRACT
\end{center}

We considered the quark-meson coupling model for a nucleon. The
model describes a nucleon as an MIT bag, in which quarks are
coupled to the scalar and the vector mesons. A set of coupled
equations for the quark and the meson fields are obtained and are solved
in a self-consistent manner. 
We show that the mass of 
a dressed MIT bag interacting with $\sigma$- and $\omega$-meson fields
differs considerably from the mass of the free MIT bag. 
The effects of the density-dependent bag constant are investigated.
The results of our calculations imply that the self-energy of the
bag in the quark-meson coupling model is significant and needs to
be considered in doing the calculations for nuclear matter or
finite nuclei.

\par
\vspace{2ex}

\section{Introduction}
For more than a decade the success of  
quantum hadrodynamics \cite{SW}
has been rather impressive 
in describing the bulk properties of nuclear
matter as well as the properties of finite nuclei.
The model is rather simple with only a few parameters, and yet it has been
successfully applied to a great number of problems for nuclear matter and
nuclear structure.  In this model
the relevant degrees of freedom are nucleons and mesons, but
the nucleons having a composite structure are treated as Dirac
particles.
Several years ago a model
to remedy this problem was proposed by Guichon \cite{Guichon}.
He proposed a quark-meson coupling (QMC) model,
in which quarks and mesons are explicitly dealt with.
The model describes nuclear matter as
non-overlapping, static, spherical
MIT bags \cite{MIT,Bagmod} interacting through the
self-consistent exchange of scalar ($\sigma$) and 
vector ($\omega$) mesons in the
mean-field approximation. 
Using this model, he 
investigated the direct quark degrees of freedom in nuclear matter.
The model was
refined later to include the nucleon Fermi motion and the
center-of-mass corrections to the bag energy \cite{Fleck} and was
applied to a variety of problems [6-10].

In the MIT bag model \cite{MIT,Bagmod} the bag constant $B$ and a
phenomenological parameter $Z$ are fixed such that the nucleon
mass of 939 MeV is reproduced for some bag radius $R$. The bag
constant $B$ produces the pressure to make a bubble in the QCD
vacuum. $Z$ is to account for various corrections including the
zero-point motion. In the QMC model for nuclear matter  
the quark-$\sigma$ coupling constant
($g^q_\sigma$) and the quark-$\omega$ coupling constant
($g^q_\omega$) are additionally introduced. Usually, $B$ and $Z$ are fixed first
so that the mass of the {\it free} MIT bag becomes equivalent to the 
nucleon mass of 939 MeV for a certain bag radius.
We denote by $B^{free}$ and $Z^{free}$, respectively,
the bag constant $B$ and the parameter $Z$
for the free MIT bag. 
$B^{free}$ and $Z^{free}$ thus obtained are used 
in the QMC model calculations, and then
$g_{\sigma}^{q}$ and $g_{\omega}^{q}$ are determined 
to reproduce the binding energy per nucleon (B.E. = $-$16 MeV) 
at the saturation density ($\rho_N^0 =$ 0.17 fm$^{-3}$).

However, due to the interaction between the quarks and the
$\sigma$- and the $\omega$-mesons 
the mass of a {\it single} dressed MIT bag in free space,
when calculated with $B^{free}$ and $Z^{free}$ described above,
may be different from the nucleon mass of 939 MeV. 
Such a possible deviation of the mass of a dressed bag from the value of 
939 MeV has been neglected in 
previous QMC model calculations for nuclear matter. 
If the deviation in mass is significant, it is necessary to modify
the parameters $B^{free}$ and $Z^{free}$
before implementing them in the nuclear
matter calculations. 
In this paper we investigate this change in the nucleon mass.
In Section 2 the QMC model for a nucleon is described.
Some numerical results are presented in Section 3.
Section 4 contains a summary.

\section{Quark-meson coupling model for a nucleon}

The Lagrangian density for the MIT bag in which quark fields $\psi$
are coupled to the $\sigma$- and $\omega$- fields may be written as
\begin{eqnarray}
{\cal L } &=& \left[ \frac{i}{2} \left( \bar{\psi} \gamma^\mu \partial _\mu \psi
-(\partial _\mu \bar{\psi}) \gamma^\mu \psi \right)
-\bar{\psi} \left( g^q_\omega \gamma_\mu \omega^\mu + 
( m_q - g_\sigma^q \sigma ) \right)
\psi -B(\sigma ) \right] \theta_v \nonumber \\
& & -\frac{1}{2}\bar{\psi}\psi \Delta_s - \frac{1}{4}F_{\mu\nu}F^{\mu\nu}
+\frac{1}{2}m^2_\omega \omega_\mu \omega^\mu
+\frac{1}{2} \left[ (\partial_\mu \sigma) (\partial^\mu \sigma) 
-m_{\sigma}^2 \sigma^2 \right] ,
\label{Lag}
\end{eqnarray}
where $F_{\mu\nu} = \partial_\mu \omega_\nu - \partial_\nu \omega_\mu $,
$\theta_v$ is the step function for confining the quarks inside the bag,
and $\Delta_s$ is the $\delta$-function at the bag surface.
Here, we assume the density-dependent bag constant introduced in Ref. [10]
and express the bag constant $B(\sigma)$ as
\begin{equation}
\frac{B}{B_0} = \left[ 1- g_{\sigma} ^B \frac{4}{\delta} \frac{\sigma}{M^0 _N}
\right] ^{\delta}
\label{eq:ddb}
\end{equation}
with $M^0 _N = 939$ MeV.
We will neglect the isospin breaking and take $m_q = (m_u + m_d )/2$ hereafter.
For actual numerical calculations $m_q$ will be taken to be zero.
The mass of $\sigma$ ($\omega$) is taken as 550 (783) MeV.
From this Lagrangian density the equations of motion
for quark fields $\psi$, sigma fields $\sigma$,
and omega fields $\omega_\mu$ follow.
For $\psi$ we have 
\begin{equation}
\left[ \gamma^\mu (i \partial_\mu - g_\omega^q \omega_\mu ) - (m_q - g_\sigma^q \sigma ) \right]
\psi \theta_v = \frac{1}{2} ( 1- i \gamma^\mu n_\mu ) \psi \Delta_s,
\label{eq:psi}
\end{equation}
where $\Delta_s = - n\cdot \partial (\theta_v)$ is used.
The left hand side of Eq. (\ref{eq:psi}) gives us
the equation for the quarks inside the bag, and
the right hand side gives us  the linear boundary condition at the
bag surface. The equations for $\sigma$ and $\omega_\mu$ are, respectively,
\begin{equation}
\partial_\mu\partial^\mu \sigma + m_\sigma^2 \sigma =
\left[ g_\sigma^q \bar{\psi}\psi 
- \frac{\partial B(\sigma)}{\partial \sigma} \right] \theta_v
\label{eq:sigma}
\end{equation}
and
\begin{equation}
\partial^\nu F_{\nu\mu} + m^2_{\omega} \omega_\mu =
g_\omega^q \bar{\psi} \gamma_\mu \psi \theta_v .
\label{eq:omega}
\end{equation}

If we consider only the ground state quarks in the static spherical MIT bag
and keep only the time-component of $\omega_{\mu}$,
Eq. (\ref{eq:psi}) for $r < R$ becomes
coupled linear differential equations for $g(r)$ and $f(r)$;
\begin{eqnarray}
\frac{d f(r)}{dr} &=& -\left[ 2\frac{f(r)}{r} + E^- (r) g(r)
\right] \label{eq:psi1} \\ 
\frac{dg(r)}{dr} &=& E^+ (r) f(r),
\label{eq:psi2}
\end{eqnarray}
where $g(r)$ and $f(r)$ are the radial parts of 
the upper and the lower components of $\psi$, $i.e.,$
\begin{equation}
\psi(t,{\bf r}) \; = \; e^{-i\epsilon_q t/R} 
\left( 
\begin{array}{c}
             g(r) \\
  - i {\bf \vec{\sigma} } \cdot {\bf \hat{r}} f(r) \\
\end{array}  
\right) \frac{\chi_q }{\sqrt{4\pi}},
\label{eq:psi-tr}
\end{equation}
where ${\bf \vec{\sigma} }$ is the Pauli spin matrix and 
$\chi_q$ is the quark spinor. 
Also,
\begin{eqnarray}
E^+ (r) &=& \frac{\epsilon_q}{R} - g_{\omega}^q \omega_0 ( {\bf r} ) 
         + (m_q -g_\sigma^q \sigma ( {\bf r} ))\nonumber \\
E^- (r) &=& \frac{\epsilon_q}{R} - g_{\omega}^q \omega_0 ( {\bf r} ) 
         - (m_q -g_\sigma^q \sigma ( {\bf r} ))
\label{eq:e+-}
\end{eqnarray}
with $\sigma( {\bf r} )$ and $\omega_0 ( {\bf r} )$ being the $\sigma$- and
the time component of the $\omega$-fields.
For the ground state 
Eqs. (\ref{eq:e+-}) are the equations in 
the radial coordinate $r$ only.
The linear boundary condition from the right hand side of 
Eq. (\ref{eq:psi}), when rewritten by using Eq. (\ref{eq:psi-tr}), becomes
\begin{equation}
f \left( \frac{xr}{R} \right) = -g \left( \frac{xr}{R} \right)
\end{equation}
at the bag surface, $ r=R$, and determines the eigenvalue $x$ of the quarks.
$\epsilon_q$ is then given by $\sqrt{ x^2 + (Rm_q )^2 } = x$ for
$m_q = 0$.
$g(r)$ and $f(r)$ are normalized such that
the quark density $\rho_q ( = \bar{\psi} \gamma_0 \psi)$ 
integrated over the bag is the unity.
In the static spherical approximation
Eqs.  (\ref{eq:sigma}) and (\ref{eq:omega}) are reduced to 
\begin{equation}
(\nabla^2 - m_{\sigma}^2 ) \sigma ( {\bf r} )= 
 \left[ \frac{\partial B(\sigma)}{\partial \sigma}
  - g_{\sigma}^q (3 \rho_s ) \right] \theta (R -r)
\label{eq:sigma2}
\end{equation}
and
\begin{equation}
(\nabla^2 - m_{\omega}^2 ) \omega_0 ( {\bf r} ) = 
 - g_{\omega}^q (3 \rho_q ) \; \theta (R -r),
\label{eq:omega2}
\end{equation}
respectively,
where $\rho_s ( = \bar{\psi}\psi)$ is the scalar density
and 3 is explicitly multiplied by the densities ($\rho_s$ and $\rho_q$) here
to account for the sum over 3 quarks. 
Equations (\ref{eq:sigma2}) and (\ref{eq:omega2}),
which are nothing but the Klein-Gordon equations with source terms,
may be readily solved using the Green's function
defined by the equation
\begin{equation}
(\nabla^2 - m_{i}^2 ) G_i ({\bf r},{\bf r'})  = \delta ({\bf r}  -{\bf r'})
\;\;\;\;\;\; (i=\sigma , \omega).
\label{eq:Green}
\end{equation}
For the s-wave state $G_i ({\bf r},{\bf r'})$ may be written as \cite{Jackson}
\begin{equation}
G_i ({\bf r},{\bf r'}) = - \frac{1}{4\pi} \frac{1}{m_i r r'} \;
sinh (m_i r_< ) \; e^{-m_i r_> }.
\end{equation}
Equations (\ref{eq:psi1}), (\ref{eq:psi2}), (\ref{eq:sigma2}), and
(\ref{eq:omega2})
form  a set of coupled equations for $\psi$, $\sigma$, and
$\omega_0$, which need to be solved self-consistently.

By solving these equations  we can obtain the eigenvalue of the
quarks and the energy of the nucleon bag. One way to calculate the energy
($E_{N}$) of the nucleon 
is computing the energy-momentum tensor
$T^{\mu\nu}$ and using the relation
\begin{equation}
E_{N} = \int d^3 r \;T_{00}.
\end{equation}
$T_{00}$ may be written as
\begin{equation}
T_{00}=\left( \frac{ {\cal E}_q }{R} \bar{\psi} \gamma^0 \psi 
      + B(\sigma ) \right) \theta (R -r)
-\frac{1}{2} \left( (\nabla \omega_0 )^2 
+ (m_\omega \omega_0 )^2 \right) 
+\frac{1}{2} \left( (\nabla \sigma )^2 + (m_\sigma \sigma)^2 \right),
\label{eq:t00}
\end{equation}
where ${\cal E}_q $ is given by
\begin{equation}
\frac{ {\cal E}_q }{R}  
        = 3 \frac{\epsilon_q }{R} - \frac{Z}{R} 
        = 3 \frac{x          }{R} - \frac{Z}{R} 
\end{equation}
with the sum over 3 quarks taken into account. Correcting for spurious
center-of-mass motion in the bag, the mass of the bag at rest is
taken to be \cite{Fleck}
\begin{equation}
M_{N} = \sqrt{E_{N}^2 - \langle p^2_{c.m.} \rangle },
\label{eq:mass}
\end{equation}
where $\langle p^2_{c.m.} \rangle = \sum_{k=1}^3 \langle p_k^2
\rangle = 3 (x / R)^2 $. By calculating $M_{N}$ for each bag radius
and minimizing $M_{N}$ with respect to the bag radius $R$,
we can get the nucleon mass and the bag radius.

\section{Results}
We first present some results of our calculations of the
nucleon mass when we take the model parameters from Ref. [10]. 
There are several sets of parameters used in Ref. [10]. 
Here we only consider the cases when $R = 0.6$ fm and
$\delta = 8$.
In Fig. 1 $M_{N}$
is  plotted as a function of the bag radius $R$.
The solid  curve represents the free MIT bag mass $M_{N}^{free}$
without a coupling of the quarks with the mesons; $g_{\sigma}^q =
g_{\omega}^q = g_{\sigma}^B=0$. 
$B_0^{1/4}$ and $Z$ are taken to be 188.1 MeV
and 2.030 \cite{ST1,JJ}, respectively, which produce the minimum 
of $M_{N}^{free}$ at 939 MeV and $R= 0.6$ fm. 
Now we include the couplings between the quarks and the mesons.
We take the parameter sets used to plot Fig. 1 of Ref. [10]. 
When $g_{\sigma}^q =0$ and $\delta=8$,
we have $g_{\sigma}^B =8.45$ and $g_{\omega}^q = 3.00$
from Table 1 of Ref. [10]. 
$M_{N}$ obtained with these parameters are plotted by the
dashed curve in Fig. 1.
$M_N$ is reduced to 848 MeV with $R = 0.694$ fm.
When $g_{\sigma}^q =1.0$ and $\delta=8$, and thus   
$g_{\sigma}^B =7.26$ and $g_{\omega}^q = 3.08$ from Table 1 of Ref. [10],
we get $M_N$ as plotted by the dotted curve in Fig. 1 with 
$M_N = 834$ MeV at $R = 0.674$ fm.
Let us consider the case when $g_{\sigma}^q =2.0$ and $\delta=8$. 
Then from Table 1 of Ref. [10] 
we have $g_{\sigma}^B =5.65$ and $g_{\omega}^q = 2.85$.
$M_{N}$ obtained with these parameters are plotted by the
dash-dotted curve in Fig. 1 with 
$M_N = 830$ MeV and $R = 0.659$ fm.
Fig. 1 of Ref. [10] 
shows that for these three different
sets of parameters the nuclear matter binding energies calculated by    
the QMC model are the same particularly near the saturation
density. However, the present calculations show that the
nucleon masses calculated with these parameter sets are 
considerably different from each other, 
all being far from $M_N ^{free} = 939$ MeV.
The couplings induce large attraction and reduce the nucleon mass 
by about 100 MeV.
Fig. 1 also shows that the attraction is not uniform
with respect to the bag radius. There is more attraction at larger bag radii,
and as a result the bag radius increases significantly.
\centerline{\epsfysize=9.5cm \epsffile{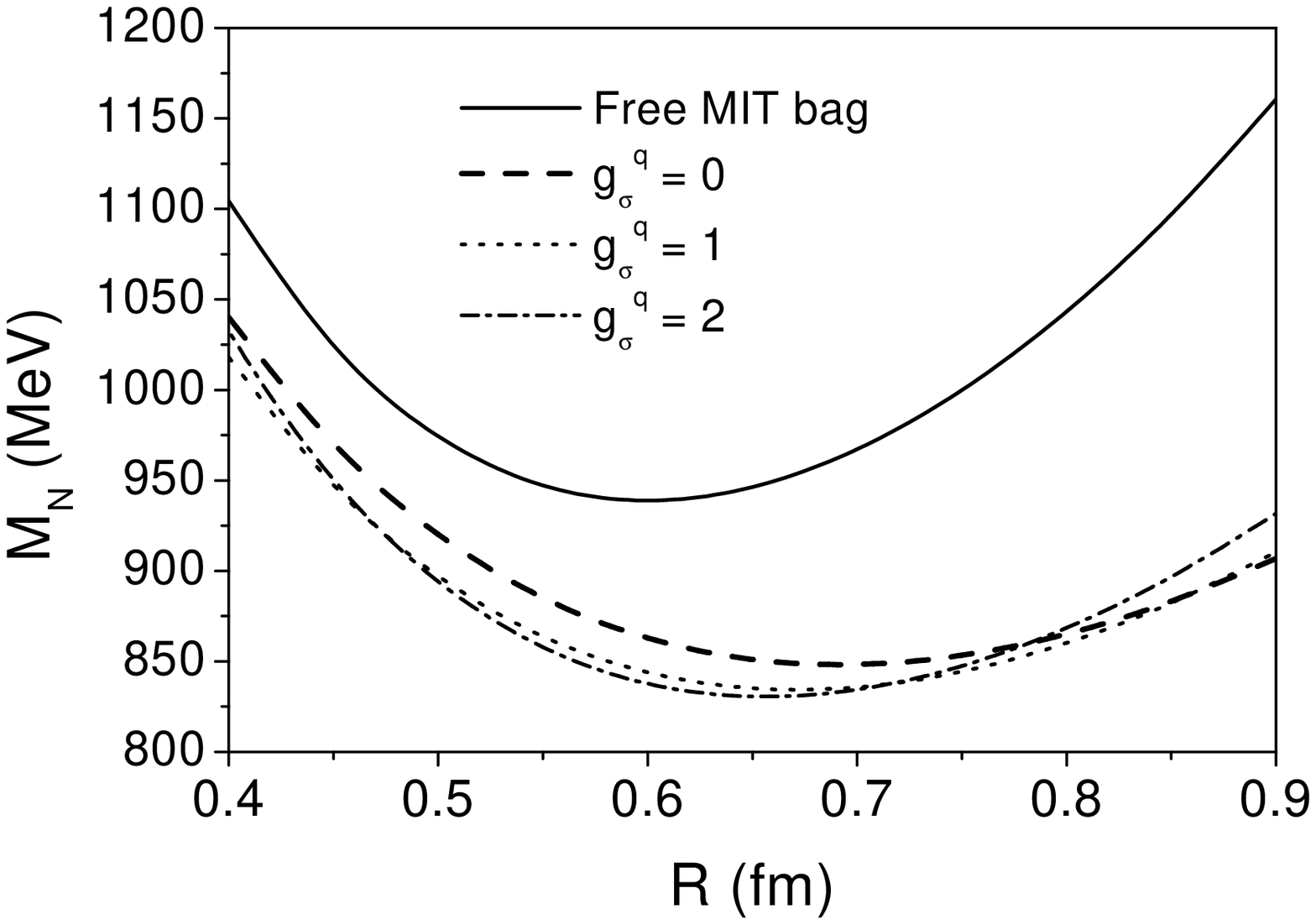}}
\vspace{-1.7cm}
\begin{center}
\vspace{-1.8cm}
\parbox{12cm}{ {\footnotesize Fig. 1 :
Nucleon mass $M_N$ plotted as a function of the bag radius $R$ for different 
coupling constants, $g_{\sigma}^q$, $g_{\omega}^q$ and $g_{\sigma}^B$.
$B_0^{1/4} = 188.1$ MeV and $Z$ = 2.030 are used for all calculations. 
For the dashed curve, $g_{\sigma}^q = 0$, $g_{\omega}^q = 3.00$ and
$g_{\sigma}^B = 8.45$. For the dotted curve, $g_{\sigma}^q = 1$,
$g_{\omega}^q = 3.08$ and $g_{\sigma}^B = 7.26$. For the dash-dotted
curve, $g_{\sigma}^q = 2$, $g_{\omega}^q = 2.85$ and $g_{\sigma}^B = 5.65$.}}
\end{center}
\vskip10pt

To see how these results come about we consider some simple cases.
Let's go back to the free MIT bag case; 
$g_{\sigma}^q = g_{\omega}^q = g_{\sigma}^B = 0$.
We still use $B_0^{1/4} = 188.1$ MeV and $Z = 2.030$.
The free MIT bag mass is plotted again in Fig. 2 by the solid curve.
Let us introduce the coupling between the quarks and the $\omega$-meson
by use of $g_{\sigma}^q = 0$ and $g_{\omega}^q = 2$
and use a constant bag constant, {\it i.e.,} $g_{\sigma}^B = 0$.
Then the dashed curve in Fig. 2 is obtained, giving us    
$M_N = 957$ MeV at $R = 0.595$ fm due to the repulsive interaction.
When $g_{\sigma}^q = 2$ and $g_{\omega}^q = 0$,
{\it i.e.,} when there is only quark-$\sigma$ coupling,
attraction is induced and we get  
$M_N = 926$ MeV at $R = 0.601$ fm, as shown by the dotted curve.
When we take $g_{\sigma}^q = g_{\omega}^q = 2$,    
cancellation between the repulsion and the attraction 
takes place, and a small repulsion is left
as plotted by the dash-dotted curve, 
resulting in $M_N = 948$ MeV at $R = 0.598$ fm.

Now we consider the density-dependency in the bag constant
using Eq. (\ref{eq:ddb}).
We start with the case that $g_{\sigma}^q = g_{\omega}^q = 2$
and $g_{\sigma}^B = 0$, corresponding to the dash-dotted curve
in Fig. 2. This curve is plotted again in Fig. 3 by the dash-dotted curve.
($M_N ^{free}$ is also plotted in Fig. 3 by the solid curve.)
To see the effects of the density-dependency of the bag constant,
we take $g_{\sigma}^B = 1$,
keeping $g_{\sigma}^q = g_{\omega}^q = 2$.
The calculated $M_N$ is   
represented by the dashed curve. 
Reducing the bag constant by introducing the
density-dependency 
\vspace{-2.5cm}
\centerline{\epsfysize=8.3cm \epsffile{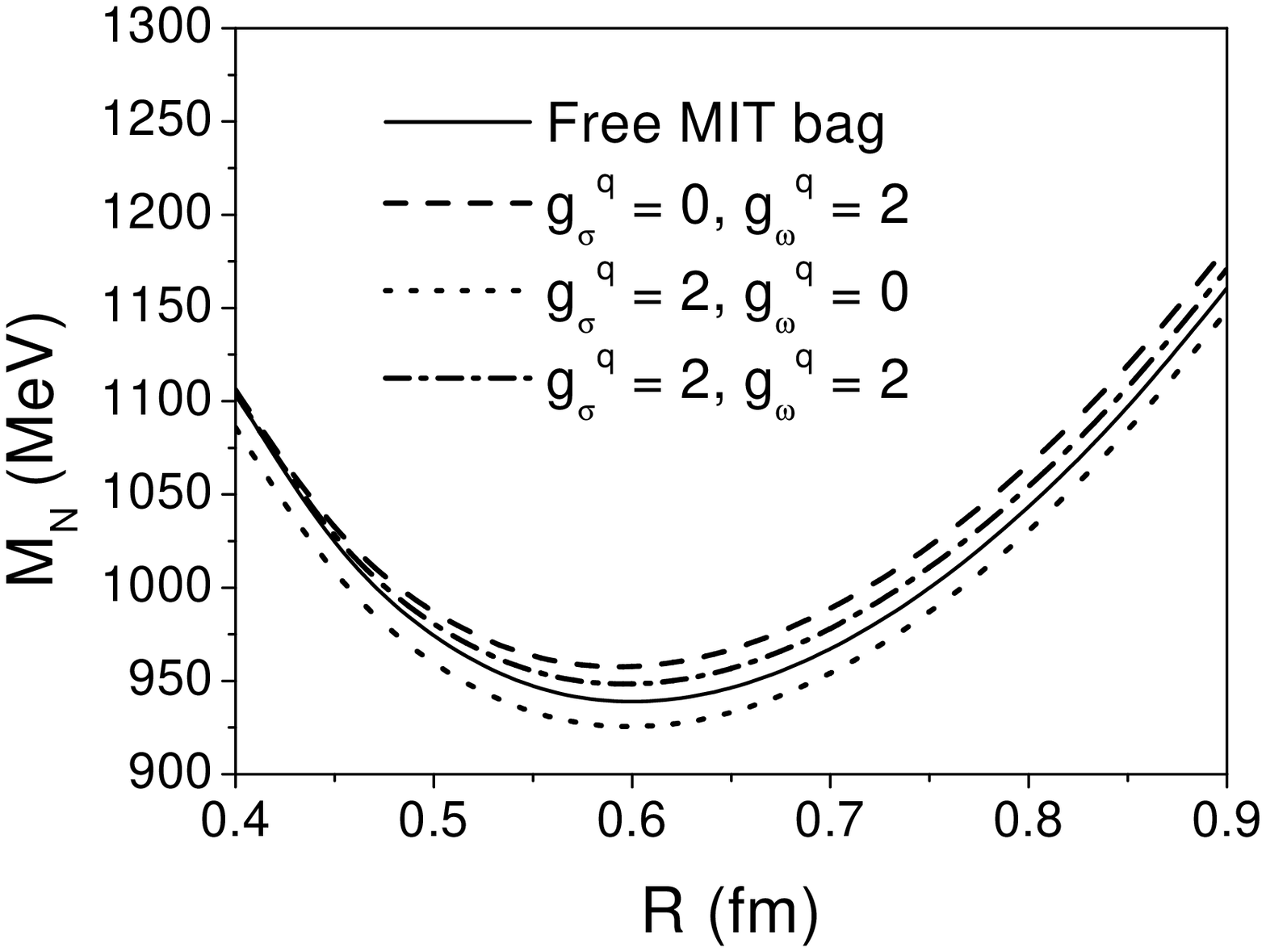}}
\begin{center}
\parbox{12cm}{{\footnotesize Fig. 2 : 
Nucleon mass $M_N$ plotted as a function of the bag radius $R$ for different 
coupling constants, $g_{\sigma}^q$, $g_{\omega}^q$ and $g_{\sigma}^B$.
$B_0^{1/4} = 188.1$ MeV and $Z$ = 2.030 are used for all calculations. 
For the dashed curve, $g_{\sigma} = 0$ and $g_{\omega} = 2$ are used.
For the dotted curve, $g_{\sigma}^q = 2$ and $g_{\omega}^q = 0$.
For the dash-dotted curve, 
$g_{\sigma}^q = g_{\omega}^q = 2$.
For all the calculations shown here $g_{\sigma}^B = 0$ are used.
}}
\end{center}
\centerline{\epsfysize=8.6cm \epsffile{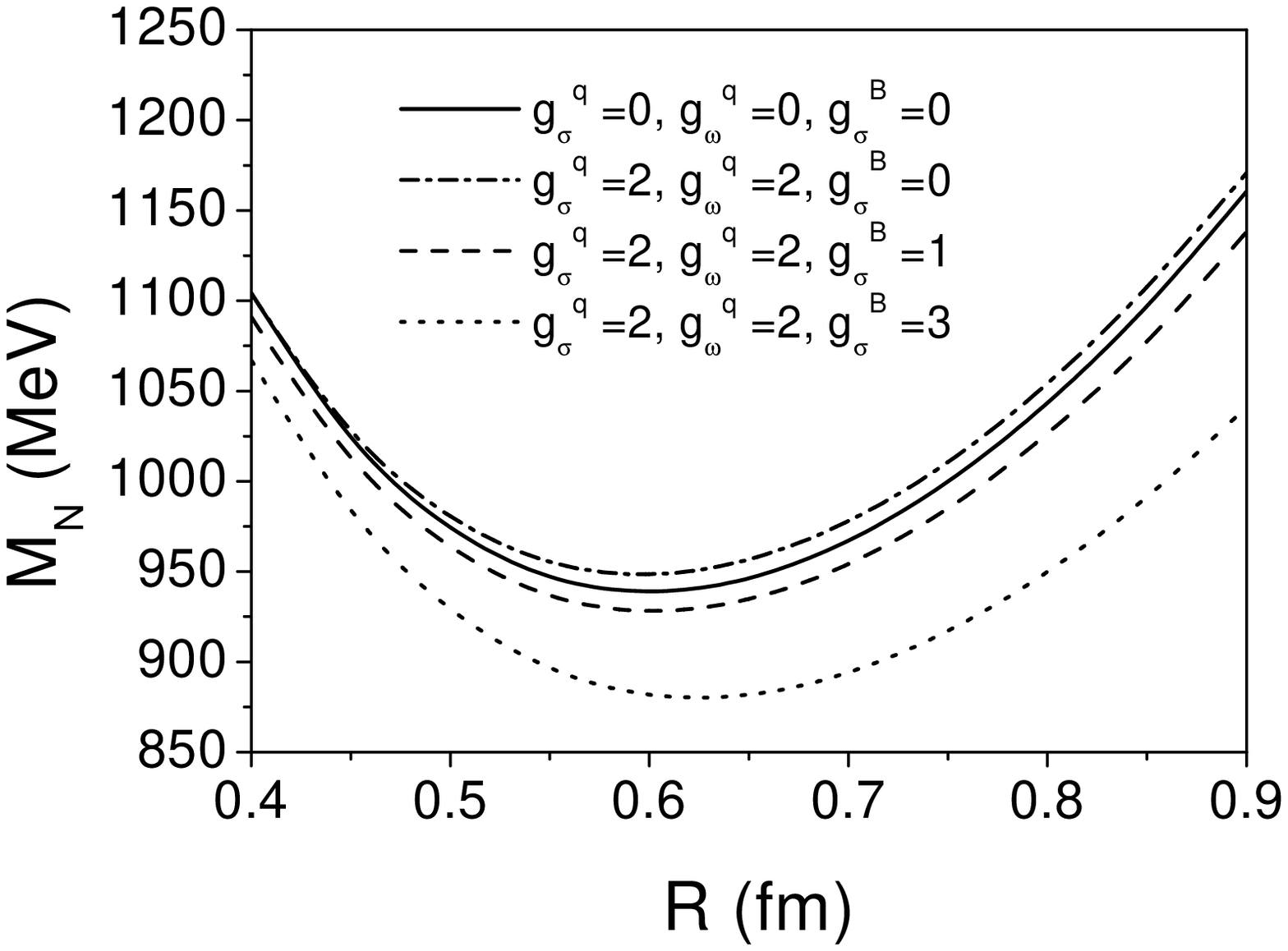}}
\vspace{-2.cm}
\begin{center}
\vspace{-1.5cm}
\parbox{12cm}{{\footnotesize Fig. 3 : Nucleon mass $M_N$ plotted 
as a function of
the bag radius with the coupling constant given in the figure.
$B_0 ^{1/4}$ and $Z$ are the same as in Fig. 2.}}
\end{center}
\vskip10pt
with $g_{\sigma}^B =1$ causes attraction.
For $g_{\sigma}^B = 3$, more attraction is induced,
as shown by 
the dotted curve in Fig. 3.
This can be seen from Eqs. (\ref{eq:sigma2}) and (\ref{eq:t00}).
The source function for the $\sigma$ field in Eq. (\ref{eq:sigma2})
increases (negatively) due to $\partial B(\sigma) / \partial \sigma$, 
so the $\sigma$ field increases, which results in enhanced attraction
in the eigenvalue. 
In Fig. 4 we show the $\sigma ({\bf r})$ field for $R = 0.6$ fm
as a function of the radial coordinate
$r$. The dotted, dashed, and  dash-dotted curves in Fig. 4 
are obtained with the corresponding 
parameters used for the dotted, dashed, and
dash-dotted curves in Fig. 3, respectively.
The figure shows that 
as $g_{\sigma}^B$ becomes larger the $\sigma$ field increases
due to the increase of the source function.
Hence, the smaller nucleon mass for the larger $g_{\sigma}^B$.
Of course, a smaller $B(\sigma)$ 
in Eq. (\ref{eq:t00}) causes a smaller mass, too.
\centerline{\epsfysize=9cm \epsffile{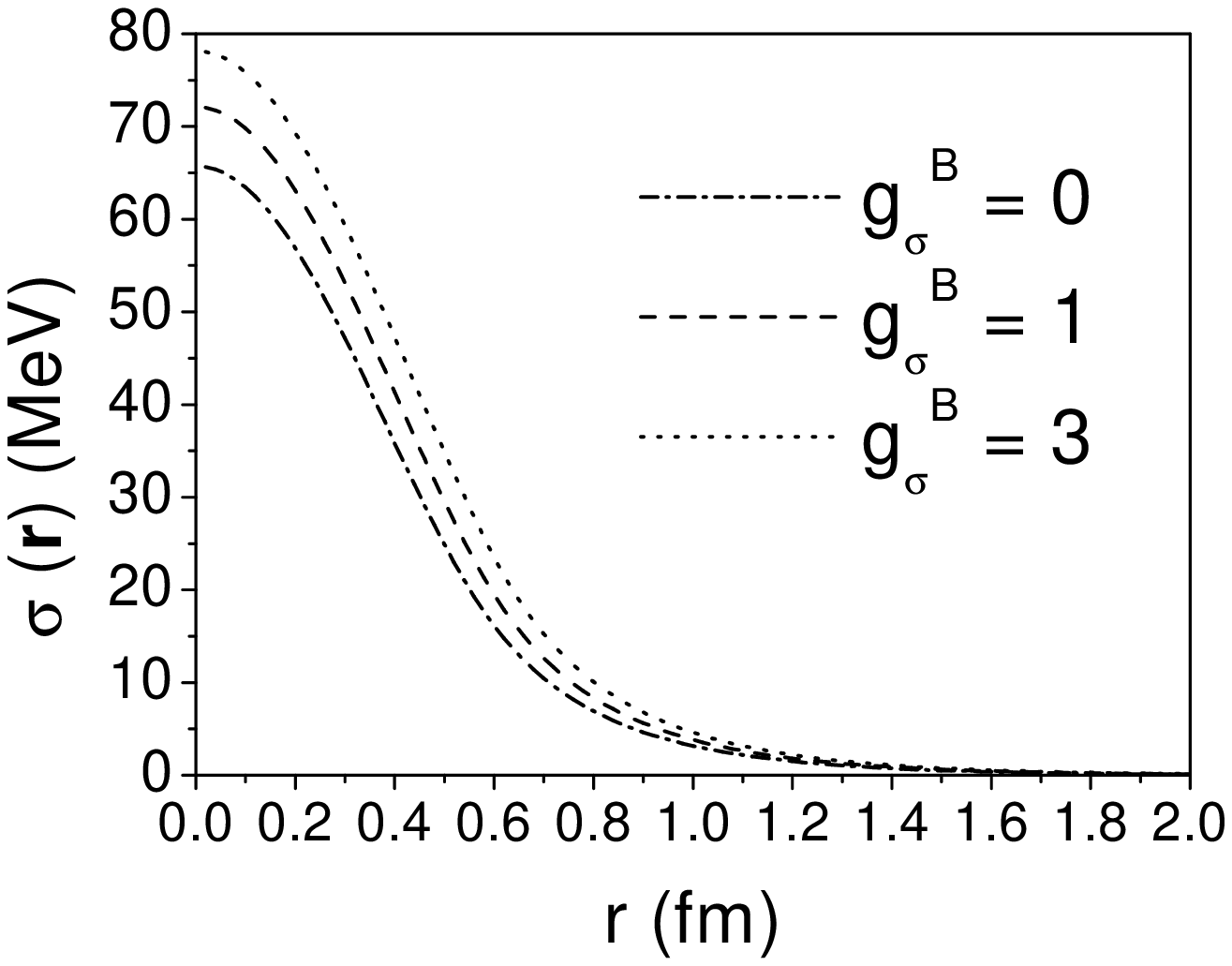}}
\vspace{-1.0cm}
\begin{center}
\vspace{-3.3cm}
\parbox{12cm}{ {\footnotesize Fig. 4 : 
$\sigma$ fields for $R = 0.6$ fm. 
The coupling constants used here for the dotted, dashed, dash-dotted
curves are the same as those for the dotted, dashed, dash-dotted
curves, respectively, in Fig. 3.}}
\end{center}
\vskip10pt

As we mentioned earlier, 
when $g_{\sigma}^B$ is large enough,
not only the nucleon mass becomes smaller,
but also the bag radius becomes larger.
When the bag constant is kept constant as in Fig. 2, the bag radii 
are essentially fixed as $R = 0.6$ fm.
Even if we change $g_{\sigma}^q$ and $g_{\omega}^q$ to some extent,
the nucleon mass curves move more or less uniformly in energy.
However, when we introduce the density-dependency in the bag constant,
the bag radius increases as shown in Figs. 1 and 3. 
The density-dependent bag constant
brings the $M_N$ curve downward in energy, but not uniformly.
There is more attraction at larger bag radii.
The reason for this can be seen from Fig. 5, where the source function 
in Eq. (\ref{eq:sigma2}) for $R = 0.6$ fm is plotted
by the solid curve.
$\partial B / \partial \sigma$ and $-g_{\sigma}^q 3 \rho_s$ are
also shown by the dotted and the dashed curves, respectively.
Due to $\partial B / \partial \sigma$ the source function does not vanish
even at large radii. 
Therefore, if the bag radius $R$ is larger, $\sigma({\bf r})$
in Eq. (\ref{eq:sigma2}) becomes greater for large radii.
This is shown in Fig. 6, where $\sigma ({\bf r})$ is plotted
for $R =0.6$ and $0.8$ fm.
$\sigma ({\bf r})$ for $R=0.8$ fm is greater than
$\sigma ({\bf r})$ for $R=0.6$ fm at large radial region,
which is the significant radial region.
As a result, there is more attraction 
at large bag radii.

\begin{center}
\centerline{\epsfysize=9cm \epsffile{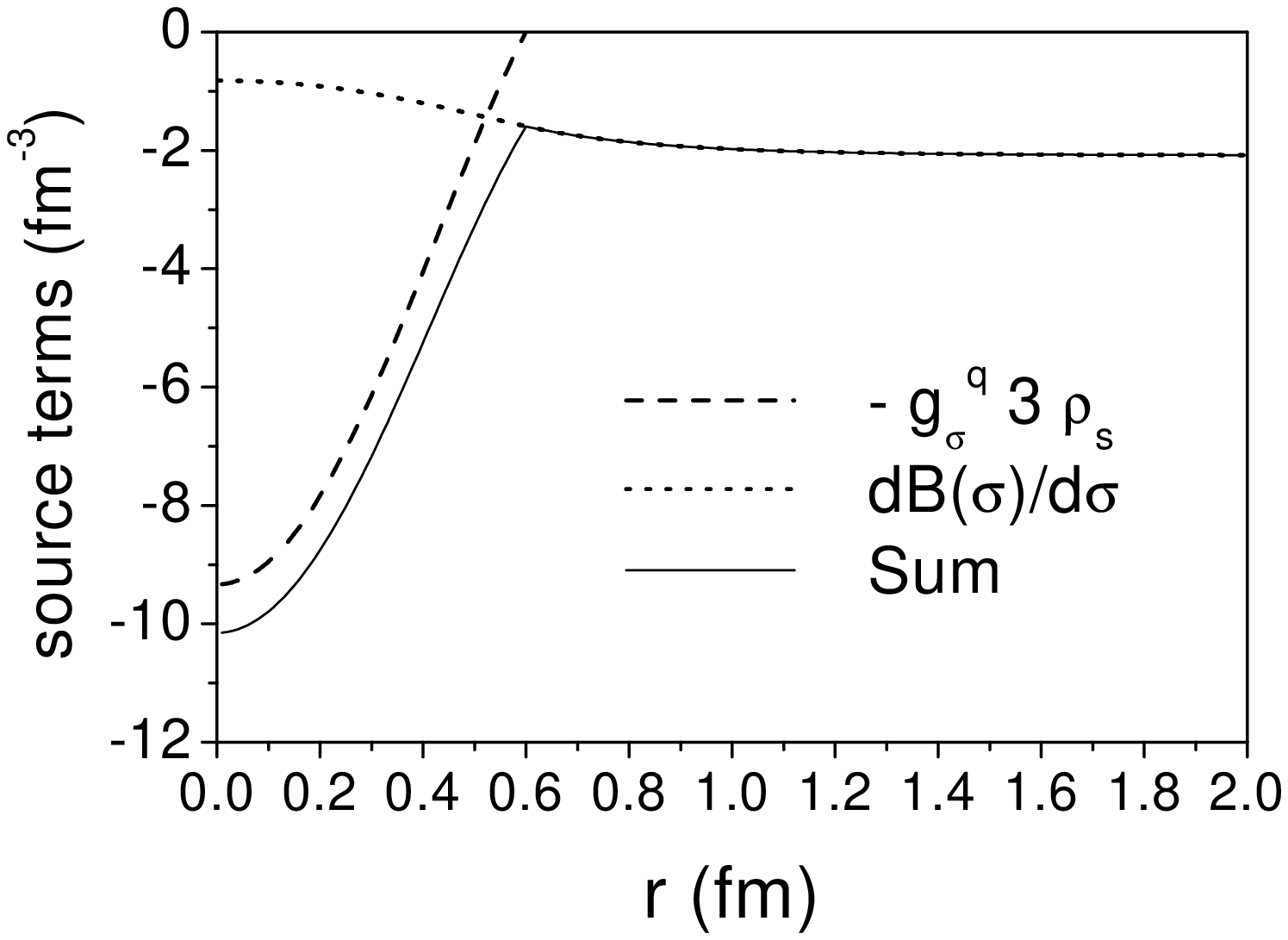}}
\vspace{-0.5cm}
\vspace{-2.0cm}
\parbox{12cm}{ {\footnotesize Fig. 5 : 
The source function is plotted by the solid curve.
$\partial B / \partial \sigma$ and $-g_{\sigma}^q 3 \rho_s$
are plotted by the dotted and the dashed curves, respectively.
The coupling constants used are $g_{\sigma}^q = g_{\omega}^q = 2$
and $g_{\sigma}^B =3$.}}
\end{center}
\vskip10pt
\vspace{-0.3cm}
\centerline{\epsfysize=8cm \epsffile{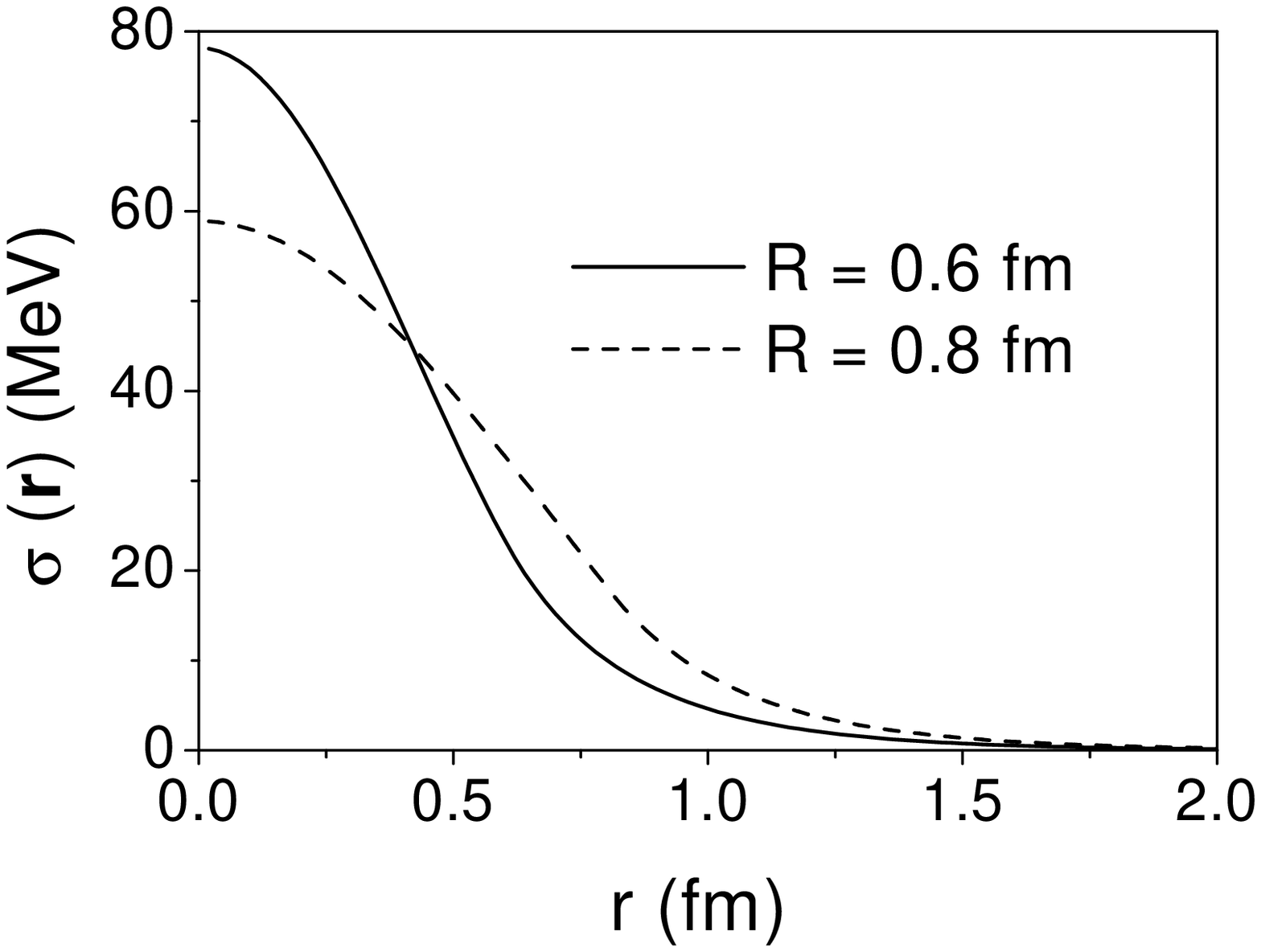}}
\vspace{-1.5cm}
\begin{center}
\vspace{-1.3cm}
\parbox{12cm}{ {\footnotesize Fig. 6 : 
$\sigma ({\bf r})$ is plotted with respect to the radial coordinate $r$
by the solid and the dashed curves, respectively, for $R=0.6$ and 0.8 fm.
The coupling constants used are $g_{\sigma}^q = g_{\omega}^q = 2$
and $g_{\sigma}^B =3$.}}
\end{center}
\vskip8pt

We made similar studies for other bag radii 
ranging from 0.6 to 1.0 fm and obtained similar results.
Our
results show that the change in the bag
mass due to 
the  coupling of the quarks with the mesons is not negligible. 
Such self-energy effects 
on the mass due to the meson coupling need to be
taken into account in choosing the parameters for 
the calculation of nuclear matter properties.
The density-dependent bag constant can not only
change the nucleon mass but also
shift the bag radius.

\section{Summary}
We have applied the QMC model to a single nucleon.
Recently the model has been often used to investigate explicit quark
degrees of freedom in describing the nuclear matter.
However, in the previous calculations 
the change in the mass of the
bag due to the self-energy was ignored.
Our calculations suggest that this change in the nucleon mass is not
negligible, and thus the model parameters need to be modified
to take this effect into consideration when they are
used in nuclear matter calculations.

\section{Acknowledgement}

This work was supported by the Natural Sciences and Engineering Research
Council of Canada.
SWH acknowledges kind hospitality at TRIUMF and 
was partially supported by Sungkyun Faculty Research Fund
and KOSEF Korea-Japan International Collaboration Program (986-0200-003-2).

\end{document}